\documentstyle[aps,psfig]{revtex}
\begin{document}

\title{
Centrality dependence of \mbox{$\pi^-$} Production and Stopping in
p-A Collisions at 18 GeV/c
}

%
%
\author{The BNL E910 Collaboration}
\author{
	I.~Chemakin$^{2}$,
	V.~Cianciolo$^{7,8}$,
	B.A.~Cole$^{2}$,
	R.~Fernow$^{1}$,
	A.~Frawley$^{3}$,
	M.~Gilkes$^{9}$,
	S.~Gushue$^{1}$,
	E.P.~Hartouni$^{7}$,
	H.~Hiejima$^{2}$,
	M.~Justice$^{5}$,
	J.H.~Kang$^{11}$,
	H.~Kirk$^{1}$,
	N.~Maeda$^{3}$,
	R.L.~McGrath$^{9}$,
	S.~Mioduszewski$^{10}$,
	D.~Morrison$^{10,1}$,
	M.~Moulson$^{2}$,
	M.N.~Namboodiri$^{7}$,
	G.~Rai$^{6}$,
	K.~Read$^{10}$,
	L.~Remsberg$^{1}$,
	M.~Rosati$^{1,4}$,
	Y.~Shin$^{11}$,
	R.A.~Soltz$^{7}$,
	S.~Sorensen$^{10}$,
	J.~Thomas$^{7,6}$,
	Y.~Torun$^{9,1}$,
	D.~Winter$^{2}$,
	X.~Yang$^{2}$,
	W.A.~Zajc$^{2}$,
	and Y.~Zhang$^{2}$,
	}

%
%
\bigskip
\address{
$^{1}$ Brookhaven National Laboratory, Upton, NY 11973\\
$^{2}$ Columbia University, New York, NY 10027 and Nevis Laboratories, Irvington, NY 10533\\
$^{3}$ Florida State University, Tallahassee, FL 32306, \\
$^{4}$ Iowa State University, Ames, IA 50010, $^{5}$ Kent State University, Kent, OH 44242\\
$^{6}$ Nuclear Science Division, Lawrence Berkeley National Laboratory, Berkeley, CA 94720 \\
$^{7}$ Lawrence Livermore National Laboratory, Livermore, CA 94550 \\
$^{8}$ Oak Ridge National Laboratory, Oak Ridge, TN 37831\\
$^{9}$ State University of New York at Stonybrook, Stonybrook, NY 11794\\
$^{10}$ University of Tennessee, Knoxville, TN 37996, $^{11}$ Yonsei University, Seoul 120-749, Korea \\
}

\date{\today}
\maketitle
\bigskip
\begin{abstract}
First results are presented from BNL experiment E910 on pion
production and stopping in proton-Be, Cu, and Au collisions at a beam
momentum of 18~GeV/c. We characterize the centrality of the collisions
using the measured number of ``grey'' tracks, \mbox{$N_{\rm grey}$},
and a derived quantity, \mbox{$\nu$}, the number of inelastic
nucleon-nucleon scatterings suffered by the projectile during the
collision.  We find that for the three targets the average backward
rapidity shift of the leading proton follows a common trend versus
\mbox{$\nu$} with the projectile losing, on average, 2 units of
rapidity in the first 2-3 scatterings. The average rapidity shift
increases more slowly with subsequent scatterings reaching a maximum
of 2.5 units.  The \mbox{$\pi^-$} multiplicity measured within the
E910 acceptance saturates with increasing \mbox{$\nu$} in \mbox{p-Au}
collisions while the \mbox{$\pi^-$} multiplicity in \mbox{p-Be}
collisions increases faster with \mbox{$\nu$} than expected from the
wounded-nucleon model. Comparisons of our data with the RQMD cascade
model suggest that in very central \mbox{p-Au} collisions most of the
pions are produced near zero rapidity in the lab.
\end{abstract}

\twocolumn

For several years 
experiments at the Brookhaven National Laboratory AGS and at the CERN
SPS accelerators have studied fixed-target collisions between nuclei
to search for the formation of the quark gluon plasma. However,
interpreting results from these experiments has often been difficult 
due to our imprecise knowledge of the complicated hadronic
interactions that take place during nuclear collisions. In particular,
the physics of the initial multiple scattering of and energy release
from the incident nucleons is poorly understood, especially at the
lower energies of the AGS. Proton-nucleus (\mbox{p-A}) collisions
provide a way to study this energy deposition process, but limitations
of past experiments have reduced the utility of \mbox{p-A} data in understanding
the physics of heavy ion colllisions. Past measurements
were obtained using either small-acceptance spectrometers
\cite{Abb92:Measurement,All70:Highenergy,Eic72:Particle} capable only of
inclusive measurements 
or low-rate ``visual'' detectors (emulsion, bubble chambers or streamer
chambers) \cite{Abe88:Leading,Bri89:PA,Dem84:PA} with 
limited particle identification capabilities. 

%
%
%
AGS Experiment 910 \cite{P910Short}
was designed to address the deficiencies in existing \mbox{p-A} data-sets by
performing semi-inclusive measurements of \mbox{p-A} collisions at AGS
energies in a large-acceptance, moderate-rate detector. In this letter
we present new measurements of two basic properties of \mbox{p-A}
collisions, the backward rapidity shift of the projectile
(``stopping'') and the subsequent production of negative pions
at a beam momentum of 18~GeV/c. Both studies are performed as 
a function of the multiplicity of ``grey'' protons and deuterons
(\mbox{$N_{\rm grey}$}), and 
we use \mbox{$N_{\rm grey}$} to estimate \mbox{$\nu$}, the number
of discrete inelastic nucleon-nucleon (\mbox{N-N}) scatterings of the
projectile in the target nucleus \cite{e910slowp}.   

The E910 spectrometer was designed to have both large geometric
acceptance and extensive particle identification coverage.
Its main component is the EOS TPC \cite{Rai90:TPC}, a large volume (${\rm
150~cm \times 100~cm \times 75~cm}$) TPC read out on a $128 \times
120$ pad segmented cathode plane \cite{Rai92}. The experiment was
staged in the Multi-Particle Spectrometer (MPS) facility at the AGS
with the TPC located at the center of the 6~m MPS ``C'' magnet which 
was operated with a nominal central field of 0.5~T. 
The TPC was oriented with its long axis 
parallel to the incident beam and its short axis (drift direction)
parallel to the primary component of the magnetic field. It was
operated with P10 gas at atmospheric pressure and a drift field of
120~V/cm.  
Particle identification in E910 is provided by \mbox{$dE/dx$} measurements in
the TPC, a 32-slat time-of-flight (TOF) wall and a 96-mirror segmented
gas Cherenkov counter operated with Freon 114 at atmospheric
pressure. Additional details on the E910 apparatus may be found in 
\cite{e910slowp}.  


The data presented in this paper were obtained from a secondary proton
beam of typical intensity $3 \times 10^{4} s^{-1}$ normally incident
on Be, Cu, and Au targets of thickness 3.9, 4.2, and 3.4 ${\rm g\,
cm^{-2}}$, respectively. The secondary beam was measured to have an
actual mean momentum of $17.5\pm 0.2{\rm \, (syst)~GeV/c}$ with a
fractional spread of 1\%. Pions and kaons in the beam were rejected
using three beam-line Cherenkov counters. Beam protons were detected
in two scintillator counters and tracked in two MWPC's upstream of the
target while halo was rejected using two ``veto'' counters. The data
presented here were triggered on the presence of a beam proton
upstream of the target and its subsequent absence in a downstream
``Bull's- Eye'' (BE) counter with 0.13 msr aperture centered on the
beam. 
Using pulse-height information from the TPC we calculated
the average \mbox{$dE/dx$} of all reconstructed particles using a 70\%
``truncation'' cut, obtaining a resolution $\sigma/\langle dE/dx
\rangle = 8\%$ for ``typical'' track lengths. This resolution allows
3-$\sigma$ or better $\pi$-p separation up to a momentum of 1.2~GeV/c. 
%
%
For the leading particle analysis we identify and reject \mbox{$\pi^+$}'s with
$p > 3.5~{\rm GeV/c}$ using the Cherenkov counter with an estimated 90\%
efficiency, and in the interval $1.2 < p < 2.7~{\rm GeV/c}$ we reject
pions at the 3-$\sigma$ level using time-of-flight. 
Due to the large acceptance of the TPC and the particle identification
detectors, we have essentially complete acceptance for protons down to
$y \sim 0.3$. For the multplicity analysis, we identify
\mbox{$\pi^-$}'s by requiring them to have a measured \mbox{$dE/dx$} within  
2.6-$\sigma$ of the nominal pion \mbox{$dE/dx$} at any given
momentum. We ignore the presence of \mbox{K$^-$}'s and
\mbox{$\bar{p}$}'s which contaminate the \mbox{$\pi^-$} multiplicities
by at most 2\%. A larger contamination results from electrons produced
primarily by photon conversions in the target. In this analysis we
implemented an \mbox{$e^+e^-$} recontruction algorithm that removes
50\% of the conversion electrons.  Fig.~\ref{fig:piaccept} shows
E910's \mbox{$\pi^-$} acceptance as a function of rapidity ($y$) and
\mbox{$p_\perp$}. The acceptance has a relatively sharp cutoff at
$y=0.5$. We make no correction to our multiplicities for 
localized tracking and particle-identification inefficiencies for
pions within our geometric aperture but we estimate these losses to be
be typically of order 3-4\%.

In the final stages of analysis we apply cuts to all reconstructed
events to remove off-target, upstream, and downstream interactions. We
require ``interactions'' to have either two charged particles in the final
state or a single charged particle with $p < 12$~GeV/c and
$\mbox{$p_\perp$} > {\rm 0.06 \, GeV/c}$.  A study of ``target-out'' data
indicates that after these cuts, approximately $ 3\%$ of our accepted
events result from off-target interactions. After all of our cuts, the
data sample for this letter contains 101k, 49k, and 36k 
events for the Be, Cu, and Au targets, respectively. 

As noted above, we characterize the centrality of \mbox{p-A} collisions using
\mbox{$N_{\rm grey}$} which has been shown to
be related to \mbox{$\nu$}, the number of inelastic \mbox{N-N} scatterings suffered
by the projectile in traversing the target
\cite{And78:PAnu,Heg81:PAnu,Heg82:PAnu,Dem84:PA}. The grey tracks 
result from the recoil nucleon ``shower'' induced
by the scatterings of the projectile in the nucleus
\cite{And78:PAnu,Heg81:PAnu,Heg82:PAnu}. 
%
%
Due to intrinsic fluctuations in this recoil shower
and the acceptance of the experiment, the dependence of \mbox{$\nu$} on
\mbox{$N_{\rm grey}$} is ``statistical'' and is not valid event-by-event.
Nonetheless, the correlation between \mbox{$\nu$} and \mbox{$N_{\rm grey}$} is
sufficiently strong that an extracted average \mbox{$\nu$}
for a given bin in 
\mbox{$N_{\rm grey}$} (\mbox{$\bar{\nu} (N_{\rm grey})$}) provides an
effective measure of collision 
centrality. Different techniques for calculating this quantity
\cite{And78:PAnu,Heg82:PAnu} have been used by other experiments in
the past \cite{Bri89:PA,Dem84:PA,Alb93}, and these techniques
plus a new variant have been studied in 
the context of E910 \cite{e910slowp}. We define \mbox{$N_{\rm grey}$} 
to be the sum of the multiplicity of protons in the momentum
range $[0.25,1.2]~{\rm GeV/c}$ and the multiplicity of deuterons in the
range $[0.5,2.4]~{\rm GeV/c}$ \cite{e910slowp}. The lower
limits are set at the approximate momentum above which secondary
recoil nucleons dominate over nuclear fragmentation
products~\cite{e910slowp}. The upper limits are determined by our
ability to uniquely identify protons and deuterons using only \mbox{$dE/dx$}. 
We use both \mbox{$N_{\rm grey}$} and \mbox{$\bar{\nu} (N_{\rm grey})$} for centrality measurements in
the analysis described below. 

The dynamics of secondary particle production in \mbox{p-A}
collisions must depend on the energy loss of the projectile and 
we have chosen to study this problem using the ``leading-particle''
technique \cite{Abe88:Leading,Too87} where we assume that the
largest rapidity (identified) proton in an event carries the baryon
number of the projectile. We quantify the projectile stopping
by the the backward rapidity shift,  $\mbox{$\Delta y$} = y_{\rm 
beam} - \mbox{$y_{\rm lead}$}$. 
%
%
We note that for \mbox{$p_{\rm beam} \gg \sqrt{m^2 + p_\perp^2}$,}
$y_{\rm beam} \approx \ln{(2 p_{\rm beam}/m)}$. 
Using this approximation the final momentum of the leading proton is 
\begin{equation}
p' \approx p_{\rm beam} \, e^{-\Delta y}.
\label{eq:deltayp}
\end{equation}
The assumption that the leading proton is a direct fragment of the 
projectile can be violated in variety of ways, but the most
significant contribution comes from charge-exchange events with high
momentum neutrons and a low-momentum (leading) protons. We have
attempted to remove such events using a variety of 
cuts, the most effective of which appears to be a requirement that the
\mbox{$\sum p_\perp$} in an event satisfies 
\begin{equation}
\mbox{$\sum p_\perp$} > 1.5\, {\rm GeV/c} \, \cdot (\mbox{$\Delta y$} - 1.1 ), \,\, {\rm for}\, \mbox{$\Delta y$} > 1.1.
\label{eq:chgexchcut}
\end{equation}
This cut requires that large \mbox{$\Delta y$} events have a corresponding
fraction of the released energy observed in the transverse motion of
the particles in the final state, and it has the largest
effect for small \mbox{$N_{\rm grey}$} since these interactions
typically have small ``true'' \mbox{$\Delta y$}'s but large observed
\mbox{$\Delta y$}'s. The parameters were chosen based on studies of
both data and the RQMD cascade model \cite{Sor95}. 

Figure~\ref{fig:deltay} shows \mbox{$\langle \Delta y \rangle$} as a
function of both \mbox{$N_{\rm grey}$} and \mbox{$\nu$} for the three
targets. We observe that the data from all three targets follow the
same trend with \mbox{$\nu$} and that \mbox{$\langle \Delta y
\rangle$} rises quickly at low \mbox{$\nu$} and then reaches an
asymptotic value of $\approx 2.4$. The slight offset of the
\mbox{p-Be} data points from the other targets is well within the
systematic errors in the \mbox{$\bar{\nu} (N_{\rm 
grey})$} extraction procedure. Comparisons of various charge-exchange
cuts indicate a small systematic error, $\approx 0.2$, on the maximum
\mbox{$\langle \Delta y \rangle$} associated with the choice of
charge-exchange cut because events with large \mbox{$N_{\rm grey}$} or
\mbox{$\nu$} rarely have a high-momentum projectile fragment of either
sign. We note, however, that we can under-estimate \mbox{$\langle
\Delta y \rangle$} for the most central collisions if a recoil proton
is ejected from the target with a rapidity larger than that of the
projectile fragment. This will most often happen when the projectile
leaves the collision with a rapidity less than or comparable to that
of a typical recoil nucleon $(y \sim 1)$. Thus, we will lose
sensitivity to the stopping of the projectile for $\mbox{$\Delta y$} >
2.5$. 
%

It is important to note that our results at $\mbox{$\nu$} = 1$ are consistent
with the well-established $\mbox{$\Delta y$} \approx 1$ in \mbox{p-p}
collisions noted by Busza \cite{Bus75:HEP75}. We observe a \mbox{$\Delta y$}
for central events which is at least as large 
as that obtained at higher energies \cite{Abe88:Leading,Too87}, and which is
quantitatively consistent at mid-centrality where the bias from target
nucleons is not expected to be significant. This agreement with the
higher energy data suggests that the stopping of the projectile,
expressed in terms of \mbox{$\Delta y$} is independent of beam energy down to
AGS energies. 

An interesting implication of the fact
that $\mbox{$\Delta y$} \sim 2$ for $\mbox{$\nu$} > 2-3$ is that if
the backward rapidity shift comes from incremental energy loss of
the projectile during the multiple collision process, after the
first few collisions the projectile would have a momentum of only
2-3 GeV/c. As a result, little or no particle production would be
expected from subsequent scatterings. Future results on strange
particle production should test whether this naive stopping picture
is valid. 

To study secondary particle production we have chosen to focus on
{\mbox{$N_{\pi^-}$}, the multiplicity of \mbox{$\pi^-$}'s within the E910 acceptance. Given
our acceptance, the measured multiplicities include all \mbox{$\pi^-$} except
for those produced at or near $y = 0$. We plot
in Fig.~\ref{fig:negpion} {\mbox{$N_{\pi^-}$} as a function of \mbox{$N_{\rm grey}$} and \mbox{$\nu$}
for the three different targets. These yields have been corrected for
electrons that survive the \mbox{$e^+e^-$} rejection cuts. 
We observe that {\mbox{$N_{\pi^-}$} increases approximately proportionally 
to \mbox{$N_{\rm grey}$} and \mbox{$\nu$} for all three targets at small values of \mbox{$N_{\rm grey}$}
or \mbox{$\nu$} but that the yields from the Cu and Au targets
subsequently saturate. The effect is most dramatic for the Au target
where {\mbox{$N_{\pi^-}$} remains constant for $\mbox{$\nu$} >  3$. 

The saturation in the observed \mbox{$\pi^-$} yields in \mbox{p-Au} and 
\mbox{p-Cu} collisions is both striking and qualitatively consistent with
the remarks made above regarding the effects of the stopping of
the projectile on the production of secondary particles.
However, the interpretation of this effect also depends on our
low-rapidity acceptance cut-off. Since it is known that $\pi$
rapidity distributions in \mbox{p-A} collisions shift to lower rapidities
with increasing target size \cite{Abb92:Measurement}, the 
acceptances losses in our \mbox{$\pi^-$} yields will likely also vary with
\mbox{$N_{\rm grey}$}. We show in Fig.~\ref{fig:rqmdpi} comparisons of our \mbox{$\pi^-$} yields
with those from the RQMD before
and after the application of E910's acceptance. The RQMD data are
plotted vs \mbox{$\bar{\nu} (N_{\rm grey})$} obtained by performing
the E910 \mbox{$\bar{\nu} (N_{\rm grey})$} 
extraction procedure \cite{e910slowp} on RQMD data. The RQMD $4\pi$
{\mbox{$N_{\pi^-}$} values increase monotonically with \mbox{$\nu$}
reaching a maximum of $\approx 6$ \mbox{$\pi^-$}'s per event whereas
the multiplicity in acceptance shows good agreement with our
results. The implication of this comparison is that $\approx 2/3$ of
the \mbox{$\pi^-$}'s in the most central \mbox{p-Au} collisions lie 
below $y = 0.5$. Irrespective of RQMD, since we see no saturation in the
$N_{\pi^-}(\nu)$ for Be but similar $\Delta y(\nu)$, the saturation
in the $\pi^-$ yields in Cu and Au seems not to result from stopping
but may indicate an effect of the heavier targets. 

%
%
%

To provide further interpretation of our results, we show with a solid
line in Fig.~\ref{fig:negpion}d expectations for the $N_{\pi^-}(\nu)$
based on the wounded-nucleon (WN) model which in its most extreme form 
states that the pion yield in \mbox{p-A} interactions depends on
\mbox{$\nu$} as 
\begin{equation}
N_\pi = \frac{N_\pi^{pp}}{2} (\nu + 1).
\label{eq:woundnucl}
\end{equation}
We have used both measurements of \mbox{$\pi^-$} production in \mbox{p-p}
collisions at 12 and 24 GeV/c \cite{Blo74:Multiplicity} and 
parameterizations of the $\sqrt{s}$ dependence of \mbox{$\pi^-$} yields 
\cite{Ros75:Experimental} to estimate $N_\pi^{pp}$ at 18~GeV/c. 
We corrected these yields downward by a factor of 0.9
to account for acceptance losses in E910 \cite{wnucl_comment} 
and upward by a factor of 1.2 to
account for isospin differences.
We observe that Eq.~\ref{eq:woundnucl}
agrees with the \mbox{p-Au} data for $\nu \leq 3$ but over-predicts at
larger \mbox{$\nu$} where the \mbox{$\pi^-$} yields saturate. A surprising result from
Fig.~\ref{fig:negpion} is that the \mbox{$\pi^-$} yields in \mbox{p-Be} collisions
increase faster with \mbox{$\nu$} than suggested by the WN
prescription. In fact, a ``binary collision'' model with $N_{\pi^-}(\nu) =
N_{\pi^-}^{pp} \, \nu$, shown in Figs.~\ref{fig:negpion} with dashed
lines, does a much better job of describing our \mbox{p-Be}
yields than the WN model. We note that this surprising
conclusion may be affected by systematic errors in our
\mbox{$\bar{\nu} (N_{\rm grey})$} extraction procedures, but to make
our results consistent with the WN expectation, the most central
\mbox{p-Be} \mbox{$N_{\rm grey}$} point would have to have $\bar{\nu} \approx 3.5$ which is very unlikely. 


In summary, we have presented first results from experiment 910 at the
BNL AGS on projectile stopping in \mbox{p-A} collisions at 18~GeV/c and the
subsequent production of \mbox{$\pi^-$}'s. We observe a common trend in the
backward rapidity shift of the projectile as a function of \mbox{$\nu$} for
all three targets and we observe that the projectile loses 2 units of
rapidity in the first 2-3 collisions.  The observed \mbox{$\pi^-$}
multiplicities show no common trend among the three targets as a
function of \mbox{$\nu$}, a result that differs from previous
measurements at higher energies \cite{Bri89:PA}. The pion yields in
\mbox{p-Au} collisions saturate for $\mbox{$\nu$} > 3$ whereas the pion yields in
\mbox{p-Be} collisions increase faster with \mbox{$\nu$} than expected based on the
wounded-nucleon prescription. RQMD comparisons suggest that the
saturation of {\mbox{$N_{\pi^-}$} is due to the low-rapidity cutoff in the E910
acceptance and a dramatic backward shift of the \mbox{$\pi^-$} distributions to
$y=0$ in the most central events. Ongoing analysis of E910 pion
spectra should provide an experimental test of this inference in 
the near future. The difference in the behavior of the two observables
that we have studied with target mass is worth emphasizing. Our data
suggests that the stopping of the projectile is determined
predominantly by \mbox{$\nu$} whereas the production of secondary
pions is strongly dependent on the mass of the target. We note that
our analysis is the first to study the centrality dependence of p-Be
interactions where the multiple interactions of the incident proton
can be cleanly studied without substantial final state interactions of
secondary particles. 

%

%
%

We wish to thank Dr. R.~Hackenburg and the MPS staff, J.~Scaduto and
Dr. G.~Bunce for their support during E910 data-taking. We also thank Dr. Thomas Kirk,
BNL Associate Director for High Energy and Nuclear Physics, for his support of
our physics program.

This work has been supported by the U.S. Department of Energy 
under contracts with BNL (DE-AC02-98CH10886), Columbia University
(DE-FG02-86ER40281), ISU (DOE-FG02-92ER4069), KSU (DE-FG02-89ER40531), 
LBNL (DE-AC03-76F00098), LLNL (W-7405-ENG-48), ORNL (DE-AC05-96OR22464)
and the University of Tennessee (DE-FG02-96ER40982) and the National
Science Foundation under contract with the Florida State University
(PHY-9523974). 

\bibliography{journals,hi,nuletter}
\bibliographystyle{prsty}


\begin{figure}
\caption{The E910 acceptance for \mbox{$\pi^-$}'s identified by dE/dx
in the TPC.}  
\label{fig:piaccept}
\end{figure}

\begin{figure}
\caption{Avg. leading proton \mbox{$\Delta y$} and in-acceptance \mbox{$\pi^-$} multiplicity for p-Be, Cu, and Au collisions plotted vs
\mbox{$N_{\rm grey}$} (a, c), \mbox{$\bar{\nu} (N_{\rm grey})$} (b, d) 
Solid(dased) line in d shows expectation from
wounded-nucleon(binary-collision) model.} 
\label{fig:deltay}
\label{fig:negpion}
\end{figure}

\begin{figure}
\caption{Comparison of \mbox{$\pi^-$} yields from the RQMD cascade model with
E910 data as a function of \mbox{$\bar{\nu} (N_{\rm grey})$} for \mbox{p-Au} collisions at 18
GeV/c. closed triangles - 
RQMD in $4\pi$, diamonds - RQMD in
E910 acceptance, circles - E910 data. }
\label{fig:rqmdpi}
\end{figure}

\psfig{file=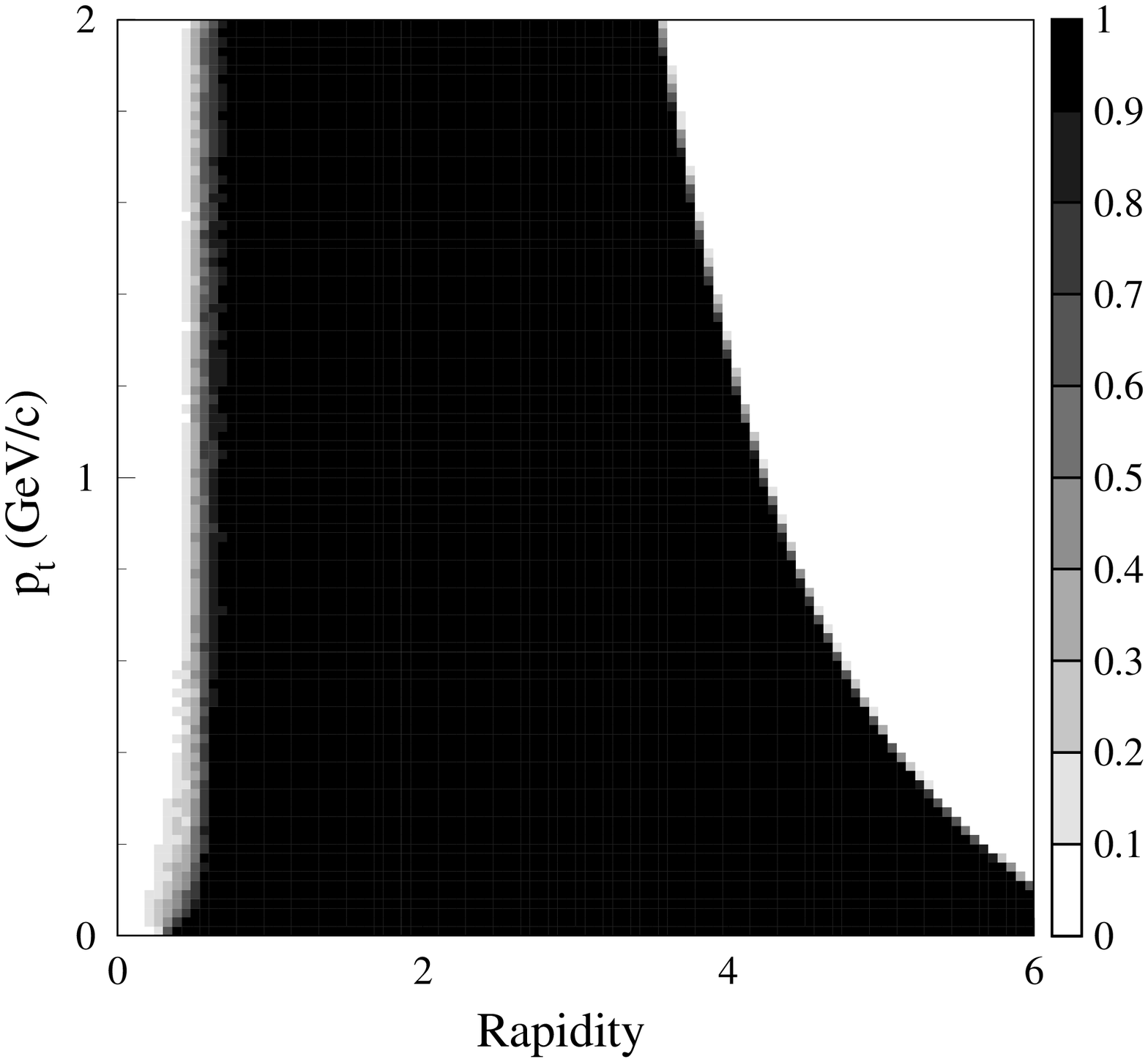,height=1.5in}

\psfig{file=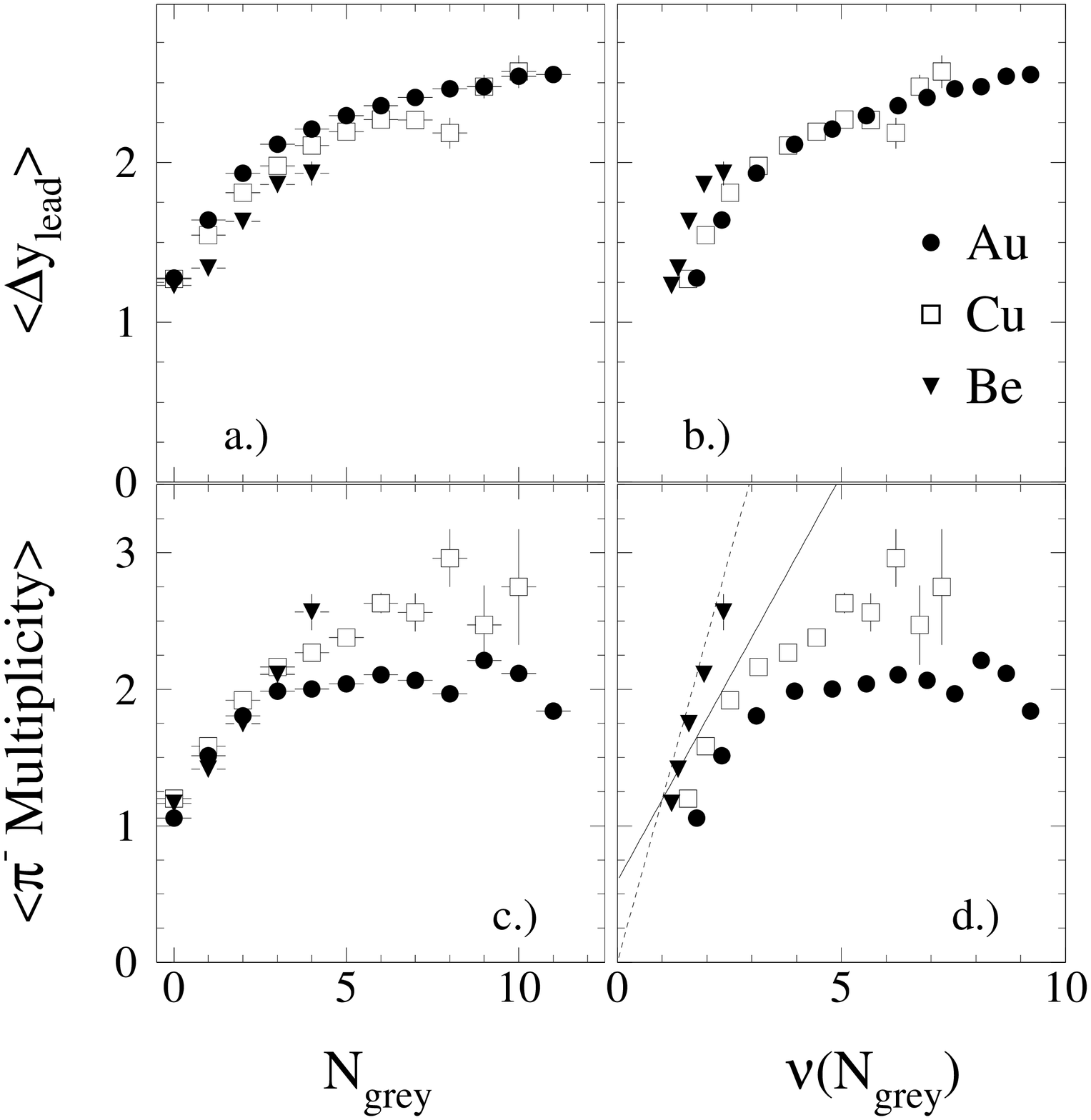,height=3.0in}

\psfig{file=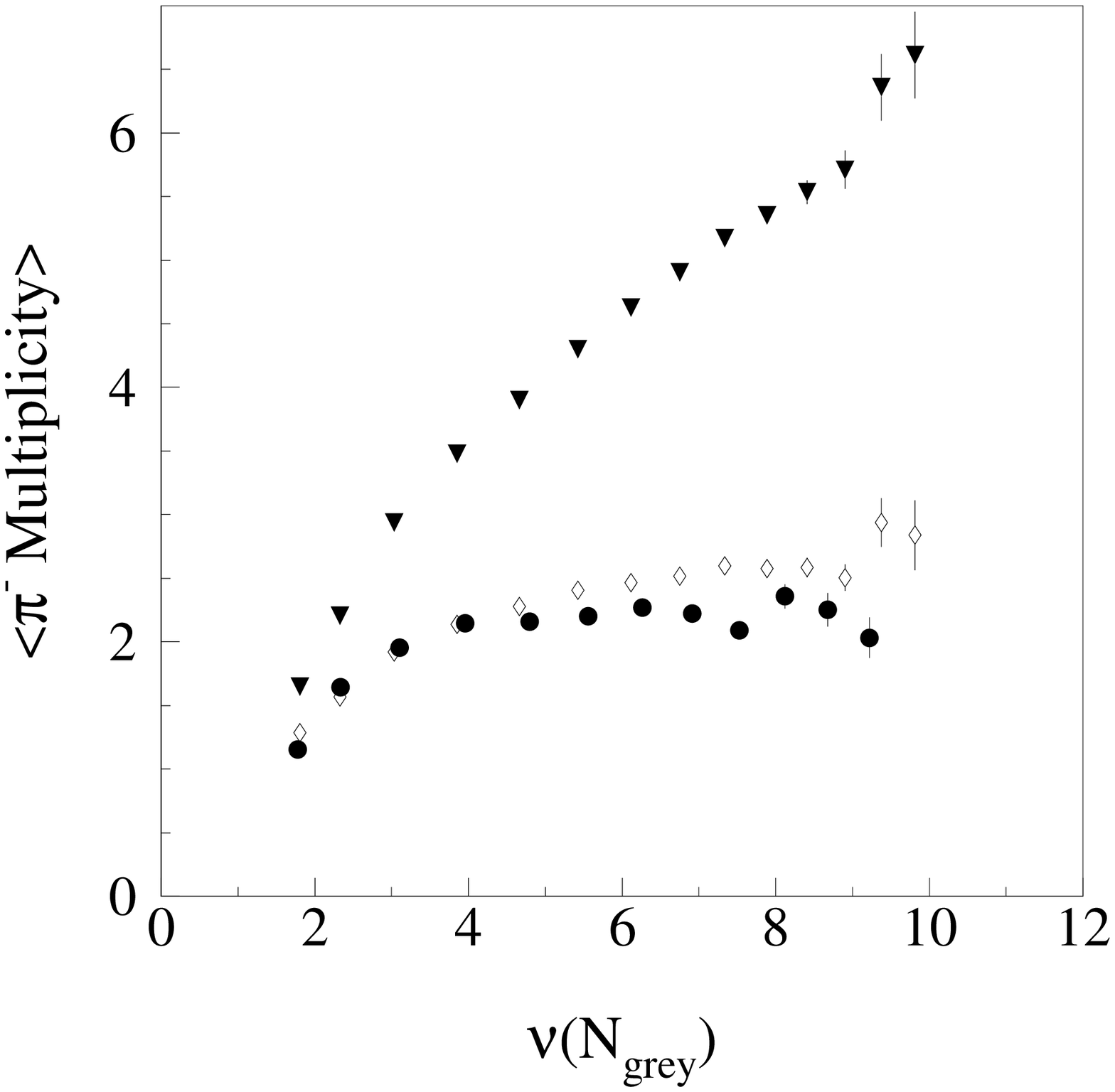,height=2.5in}

\end{document}